# SUPPORTING EDUCATION IN MATH SCIENCES WITH A LOW-BUDGET LMS SAKAI

Mimoza Petrova, Vanco Cabukovski, Roman Golubovski

**Abstract** In this paper we present the Low-budget LMS Sakai, the analysis conducted and its selection from a list of few popular Low-budget LMSs, as well as its capability in presentation of math texts and formulas. The MATHEIS e-learning system for learning mathematics and informatics and its empowering with few software agent's based features including the new developed GUPA (Graphic User Presentation Agent) software agent is emphasized. The plugins offered for Sakai allow powerful mathematics presentation capabilities, one of them being the WIRIS plugin. The paper is focused on the Sakai WIRIS plugin implementation.

## 1. INTRODUCTION

In information era we learn from different media of information: books, pictures, audio files, audio-video presentations, etc. We learn in every place and at every time sharing the knowledge with the coworkers or our mentors by pressing only one or few buttons of our mobile phones or smart devices used. Nowadays trend is to implement new technological advancements in all educational stages and in almost all educational forms. Successful accomplishment of this mission depends on quality of developed e-content (unit of information in digital form) – it attracts challenges and motivates the learners; but it is not the only key factor for success.

The e-learning is a process. This process must be delegated by some management system – Learning Management System (LMS). The key responsibilities to this system are providing learners with:
- necessary learning material (e-contents);
- ability to communicate (with each other and/or with professors/educators)
- team work;

___

- work on projects and share files.

On the other hand the LMS system must provide the professors/educators ability to:
- give assignments to the learners;
- track the learners activities;
- make quizzes and tests;
- grade learners tests;
- enroll and unenroll learners from courses.

And last but not the least the LMS system must also provide means to Government institutions for auditing, tracking the learning processes, doing some external testing, etc. - but this is out of the scope of this paper.

The LMS market is growing rapidly. E-learning training tools, administration functions, mobile accessibility, and social media communication features make LMSs attractive not only for educational organizations but in general for any kind of organization. These LMS's features make e-learning systems irreplaceable to many organizations.

Mathematics equips pupils/students with a uniquely powerful set of tools to understand and change the world. These tools include logical reasoning, problem-solving skills, and the ability to think in abstract ways. Mathematics is important in everyday life, in many forms of employment, in the science and technology, in the medicine, in the economy, in the environment and development, and in the public decision-making [2].

Mathematics educators must be prepared to empower students with the advantages technology can bring. Schools and classrooms, both real and virtual, must have educators who are equipped with technology resources and skills [12].

E-content is a very powerful tool of education. E-content is valuable to the learners and also helpful to educators of all individual instruction systems; The use of e-content is changing education [20].

There are many Learning Management Systems (LMS) like as: Moodle, Canvas, SAKAI, JoomlaLMS, aTutor, dotLRN, Claroline, Ilias, Chamilo, Dokeos, SharePointLMS, etc. [5], [9], [14]. More popular are low-budget LMS (Open Source LMS) [1], those integrating capabilities of more specific e-content creation, integration and delivery as math sciences contents are – which integrate text, formulas, special symbols, graphics, etc., import and export capabilities to specific formats and editors as well as tools based on OpenMath, MathML and similar e-content math sciences standards. OpenMath is a standard for representing mathematical objects with their semantics, allowing them to be exchanged between computer programs, stored in databases, or published on the Web. MathML is a mathematical markup language. It is an application of XML for describing mathematical notations and capturing both its structure and content. It aims at integrating mathematical formulas into worldwide web pages and other documents. MathML deals with the presentation of mathematical objects, while OpenMath is concerned with their semantic meaning or content [4].

In this paper the educational system for learning mathematics and informatics MATHEIS, its agent based extension and its adaptation with a GUPA (Graphic User Presentation Agent) agent for manipulation with math sciences e-content will be described in short. The hosting LMS system Sakai as our choice for a LMS and WIRIS equation editor, WIRIS quizzes and WIRIS calculator as a WIRIS plugin for Sakai as an essential parts of the software agent GUPA will be also described. The process of analysis for selection of the appropriate Open source LMS platform as well as the load balancing and scaling of the chosen LMS platform will be also given in this paper.

## 2. MATHEIS AGENT BASED SYSTEM IN A SAKAI LMS ENVIRONMENT

At the Faculty of Natural Sciences and Mathematics in 2007 was successfully implemented a commercial ORACLE iLearning LMS platform and the educational system for learning mathematics and informatics MATHEIS [6], [10] has been successfully integrated into this platform and migrated into Open Source LMS Sakai [7], [8]. This e-learning environment (Sakai) was chosen from a selected list of few Open Source LMSs on a bases of literature made comparisons (ATutor, dotLRN, Ilias, Moodle and Sakai) and survey we have done in order to select appropriate load balancing and scaling LMS capable to incorporate e-content math standards and tools (like OpenMath, MathML, LaTex, WIRIS, etc.). The requirement for e-content math sciences standards and tools supporting was initiated by the migration and further development of the educational system for learning mathematics and informatics MATHEIS, while the LMS balancing and scaling was initiated by the necessity of massive exploitation and cloud migration of the system in a future.

An agent based extension was developed with MATHEIS empowerring the system with the following new features: monitoring the students behaviour and interests at the system; determining student's skill level; enabling cooperative task resolution among students; enabling different views of services and content according to student's skills and requirements - adaptability; notifying students when newest tests for appropriate level are available; presenting tests to students and estimating received results; automatic update of student's levels depending on estimated results, etc.

The general idea behind this concept was to provide the students with supplemental material in support to steeper learning curves. The LMS is constantly monitoring the overall progress of all students with their semestral courses and updates the supplements ranking list accordingly. Students are also evaluated periodically through test examination and ranked accordingly. Adaptation is then implemented with an algorithm which basically suggests higher ranked supplements set to lower ranking students, and more relaxed content to higher ranking students. This adaptive aspect is implemented by the PFA subsystem given in Figure 1.

The system is able to assist adaptively in filtering educational material according to the UPA (User Profile Agent) and student's activities in communication with MATHEIS basic services recorded by the PAA (Personalized Activity Agent). The Personalized Content Viewing Agent (PCVA) is responsible for the adaptive interaction and adaptive content/course delivery. All three agents collaborate among them sharing distributed agent knowledge and learning rules [7].

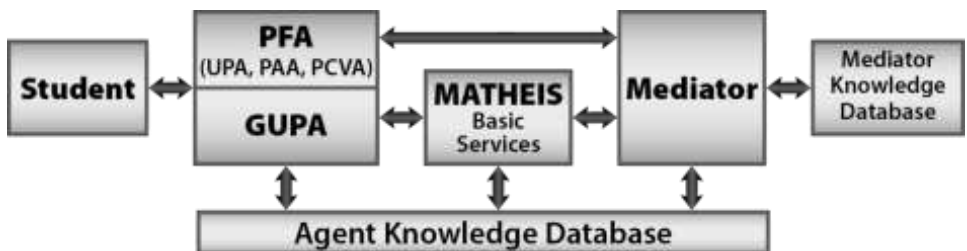

**Figure 1:** The agent-based MATHEIS

The Mediator is responsible for the student learning model, database of the student's grades, degree levels, preferences, abilities, aptitudes, etc. It communicates with the Mediator knowledge database.

The Personalized Filtering Assistant (PFA) follows the student's activities. The PFA is trained for each student to make the right content selection appropriate to the student's abilities and aptitudes.

GUPA (Graphic User Presentation Agent) is an agent developed for content editing of math sciences formulas, symbols, graphs, etc. It is based on MathOpen and MathML standards and establishes communication with some mathematical plugins as WIRIS [21] is. Further development of the GUPA is required in order to give some additional intelligence in respect to semantic meaning and reasoning of the math sciences e-content.

Cloud LMS are SAAS (software as a service) LMS, which represents a network of remote servers hosted into the cloud. As most known cloud LMSs are Google Classroom and TalentLMS [4]. Currently we are working on MATHEIS extension on a cloud incorporating all developed and implemented agent based characteristics of the system.

### 3. ANALYSIS OF SOME POPULAR OPEN SOURCE LMSS

Open Source LMSs have many benefits and many disadvantages. The main benefits are:
- avoiding vendor lock-in;
- enhanced reliability;
- license at zero cost;
- easy customization (based on needs);
- localization on languages not considered by commercial vendors.

But using Open Source LMSs has many disadvantages, like:
- raising the maintenance cost;
- there are costs for installing, support and maintenance;
- need for understanding of different open source licences.

To transfer the content from one to another LMS both systems must be SCORM compliant. SCORM (Sharable Content Object Reference Model) is e-learning standard preferable for the Open source LMS platforms increasing the LMS quality in general.

Based on some literature made comparisons we have taken following few LMS Open Source products: ATutor, dotLRN, Ilias, Moodle and Sakai as a possible choice for our solutions [5], [9], [14], [15], [16].

**ATutor [3]**

ATutor is an Open Source web-based Learning Management System (LMS) and social networking environment with accessibility and adaptability possibilities. The administrators can easily install and update the system, develop a custom themes to give a ATutor a new look, and easily extend the system functionality with the feature modules. Educators can quickly assemble, package and redistribute the web-based instructional content, easily import prepackaged content, and conduct their courses online. Students learn in an adaptive social learning environment. Its last stable version is 2.2.2. There are no relevant information and recommendations to improve the ATutor scalability. ATutor is translated in 75 languages one of which is Macedonian. Environment supported: HTTP Web Server ( Apache recommended ), PHP (Version $4.3.0^+$ for ATutor $1.6.3^+$ to 2.2.2; Version $5.0.2^+$ for ATutor $1.6.3^+$ to 2.2.2, Version 5.4 for ATutor 2.1.1. to 2.2.2), MySQL database Version $4.1.10^+$. ATutor is SCORM 1.2 export/import compliant. ATutor supports the following e-content math standards and tools: OpenMath, MathML, LaTex text editor, WIRIS equation editor and WIRIS calculator.

### dotLRN [11]

dotLRN was developed as an open source LMS platform at the Massachusetts Institute of Technology (MIT). It is used worldwide by over a half a million of users in institutions for higher education and government, non-profit organizations and US K-12 education. dotLRN is supported by dotLRN consortium a nonprofit organization that handles governance, coordination and ongoing development. Its last stable version is 2.5.0. dotLRN is translated into 40 languages. It is not translated in Macedonian yet. There are no relevant information and recommendations to improve the dotLRN scalability. Environment supported: HTTP Web server (Apache), TCL programming language,  MySQL database Version $4.1.10^+$. It is SCORM 1.2 and 1.3 export/import compliant. dotLRN supports the following e-content math standards and tools: OpenMath, MathML, LaTex text editor, WIRIS equation editor and WIRIS calculator.

### Ilias [13]

Ilias is a powerful Open Source web-based Learning Management System (LMS) for development and realizing web-based e-learning. It was developed to reduce the coasts of using new media in education and further training and to ensure the maximum level of customer influence in the software implementation. Ilias is published under the GNU (General Public License). Its last stable version is 5.1.0. Ilias is scalable by supporting load-balancing techniques and multiple database servers. Ilias is translated into 28 languages. It is not translated in Macedonian yet . Environment supported: HTTP Web server (Apache), PHP programming language,  MySQL database. It is SCORM 1.2 and 1.3 export/import compliant. Ilias supports the following e-content math standards and tools: OpenMath, MathML, LaTex text editor, WIRIS equation editor and WIRIS calculator.

### Moodle [18]

Moodle is an Open Source Course Management System (CMS), also known as a Learning Management System (LMS) or  a Virtual Learning Environment (VLE). It has become very popular among  educators around the world as a tool for online dynamic web sites creation for their students. Its last stable version is $3.1.1^+$. Moodle's design with clear separation of the application layers allows strongly scalable setups. Large sites usually separate the web server and the database onto different servers. It is possible to load balance a Moodle installation, e.g. by using more than one web server. Moodle is translated in 123 languages one of which is Macedonian. Environment supported: HTTP Web server (Apache), PHP programming language (Moodle $3.1.1^+$ requires PHP 5.4.4), databases MySQL 5.5.31, MariaDB 5.5.31, PostgreSQL 9.1, MSSQL 2008 and Oracle 10.2. It is SCORM 1.2 and 1.3 export/import compliant. Moodle supports the following e-content math standards and tools: OpenMath, MathML, LaTex text editor, WIRIS equation editor, WIRIS quizzes and WIRIS calculator.

**Sakai [19]**

Sakai is a free and open source product that is built and maintained by Sakai community. It is an online Collaboration and Learning Environment. Many people use Sakai for teaching, learning and collaboration. Many successful and big project are realized with Sakai. Its last stable version is 11.1. Sakai is a highly scalable enterprise application, Sakai's strategy for scalability will be described in a next section. Sakai is translated in 20 languages. It is not translated in Macedonian yet. In general Sakai is installable on every server running Apache Web server, Tomcat Application Server and Database Server (MySQL, Oracle, HsqlDB). It is SCORM 1.2 and 1.3 export/import compliant. Sakai supports the following e-content math standards and tools: OpenMath, MathML, LaTex text editor, WIRIS equation editor, WIRIS quizzes, WIRIS calculator, Sferyx equation editor, fMath editor and Sakai Wiki jsMath math formulae.

Sakai has developed full strategy for load balancing and scaling which will be described in the next Section.

A cloud LMS approach requires scaling LMS applications easy by load balancing user traffic, offloading the security overhead, and allowing additional server resources to be added without any user impact.

## 4. LOAD BALANCING AND SCALING IN SAKAI

Learning Management Systems have to be reliable, secure, highly affordable and available. Such mission critical applications as LMSs are, require high performance load balancing. It improves the distribution of workloads across multiple computing resources (disk drives, computers, network links, CPUs, storages, computer clusters, etc.) providing a single Internet service from multiple servers [17].

Sakai's strategy for load balancing and scaling offers topologies as a respond to growing of total number of simultaneous users. These topologies are:

- Standalone server – which is the simplest topology of installation; all Sakai's components run on a single server; could be used for standalone development and demonstrating the system usability.
- Thin client configuration consisting of two tiers - front end with Apache web server (which is optional) and Tomcat application server, and the second tier is consisting of two servers (database server and file storage which might be optional).
- Load balanced thin client consisting of three tiers – the first tier is Load Balancer, the second tier may consist of more than one server (each of these servers are Apache which could be optional and Tomcat), the third tier consists of two servers (database server and file storage which might be optional).
- Thicker client which is an application server that runs on more Tomcat instances behind an Apache instance; this topology is consisting of three tiers – the first tier is Load Balancer, the second tier may consist of more than one server; each of these servers are Apache (which could be optional) and more than one Tomcat installation, the third tier consists of two servers (database server and file storage which might be optional).

Sakai is the LMS that can be successfully implemented in all possible topologies. However ATutor, dotLRN Ilias and Moodle can be successfully in local and mixed topologies.

Final decision of which LMS to choose inclusively depends and on a user satisfaction and LMS dedication. As could be seen from previous section Sakai supports more math e-content standards and tools than the other mentioned LMSs. MATHEIS agent based system for learning mathematics and informatics was installed on Sakai environment at the Faculty of Natural Sciences and Mathematics. The developed GUPA agent is based on MathOpen and MathML standards and establishes communication with a mathematical WIRIS plugin. In a next section will be given a description of a WIRIS as a Sakai math plugin.

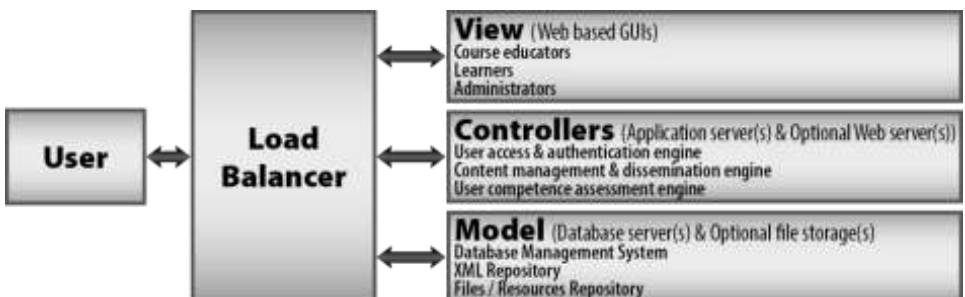

**Figure 2:** The MVC architecture of Sakai load balancing and scaling environment

In Figure 2 is given the MVC (Model-View-Controller) architecture of the Sakai load balancing and scaling environment providing access to MATHEIS e-learning system from multiple physical or virtual servers. The View and Controller layers are deployed on an application server(s) and optional Web server(s) while the Model layer represents the persistent database server(s) and optional file storage(s). A load balancer distributes network or application traffic across the servers. It increases the capacity (concurrent users) and reliability of the applications.

## 5. APPLICATION OF SAKAI IN MATH SCIENCES - WIRIS SAKAI PLUGIN

WIRIS equation editor [21] is mathematical visual (WYSIWYG) editor. WIRIS uses a large collection of icons nicely organized in thematic tabs in order to create formulas for any web content. It is JavaScript based and compatible with HTML 5. WIRIS has a special tool section for a Chemistry too.

WIRIS plugin for Sakai provides the following features:
- easy installation and integration of the WIRIS tool;
- support of different technologies (PHP, Java, ASP.NET), and different HTML editors (TinyMCE, CKEditor, etc.), or platforms (Sakai, Moodle, Joomla, etc.);
- saving the images and the formulas for reusability;
- tuning features by configuration file;
- opportunity of improvement and bug fixing.

WIRIS plugin for Sakai provides and possibility of self-testing quizzes (WIRIS quizzes), where the educators can set up a base of questions and the learners could take a practical test from home and see the level of their knowledge for a part of the material as well as WIRIS calculator (WIRIS cas) including integrals and limits calculation, function graphing in 2D or 3D and symbolic matrices manipulation, among others.

WIRIS offers solutions of very complex expressions and in the same time the process of equation is shown step by step. It allows export of the equations as an images easy importable everywhere.

WIRIS allows a self-testing. On this way the learner could see his/her own progress.

WIRIS allows to the educators easier way of audio-visual presentation creation.

In Image 3 is given an WIRIS plugin integration into Sakai. A WIRIS quizzes screen on Sakai is presented in Figure 4.

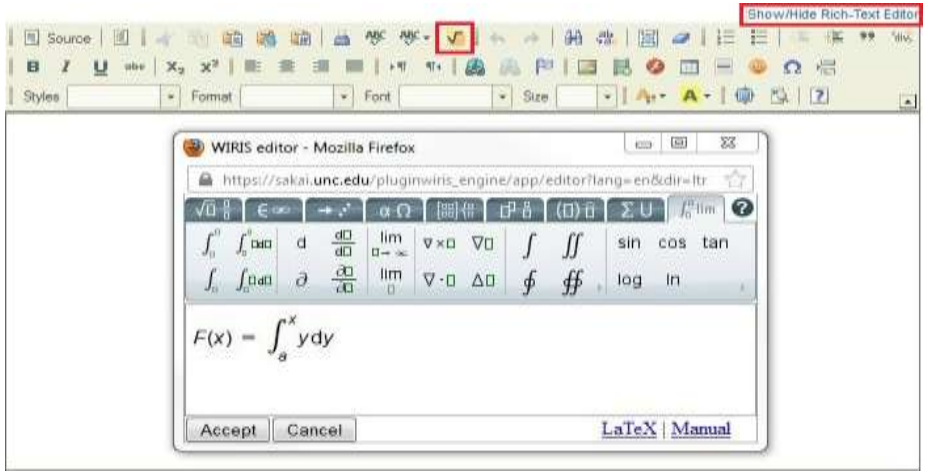

**Figure 3:** WIRIS plugin integration into Sakai

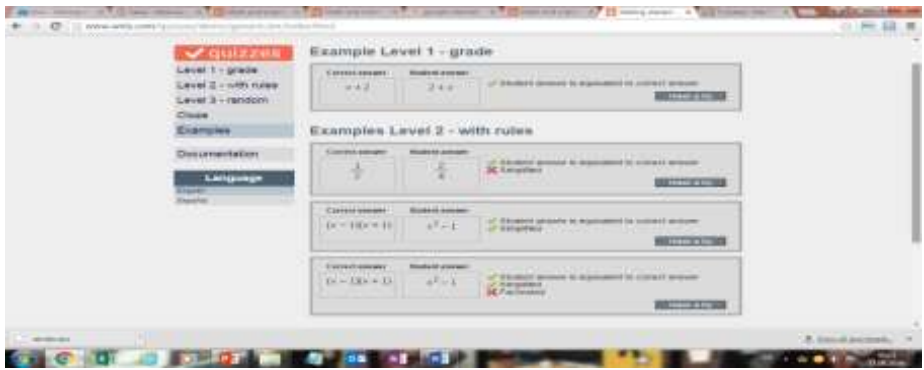

**Figure 4:** WIRIS quizzes on Sakai

## 6. SOME PROBLEMATIC ISSUES CONCERNING A WIRIS IMPLEMENTATION

It is more than evident that today e-content can bring a great potential to education, but largely this potential has not been recognized in mathematics and, more generally, in sciences. This is partly due to problems that one has in the presentation of math sciences content on the Web. Mostly the mathematical formulas and expressions are presented into a e-content as pictures only, not usable as a key words in a database searching and indexing for example. MathML and OpenMath provide solutions to this.

WIRIS equation editor can export the mathematics formulas and expressions into MathML standards but does not embed them into database indexes and make them searchable. This WIRIS shortage is solved by the GUPA software agent we have developed.

WIRIS quizzes embed the MathML formulas into their own database, use indexing and make them searchable only for themselves. In order to extend this possibility to the whole MATHEIS system we had to solve this problem by developing the GUPA software agent.

The WIRIS cas (WIRIS calculator) was integrated without problems in a whole Sakai environment including the MATHEIS e-learning system. From any MATHEIS basic service and tool the integrated WIRIS calculator can be easily used.

The complete perception of knowledge in mathematics is based on the logical abilities and reasoning of the learners. Usually they need to solve complex problems in a step by step procedure, to prove theorems which are based on more formulas and knowledge and solve the problems thinking in abstract ways. In most of these situations it is difficult for the educator to construct the wright question and find the real answer offered in WIRIS quizzes. Requesting only the final result of a solution is not always authoritative. WIRIS does not possess any intelligence or reasoning possibilities to follow up the solution of the problem step by step and provide the educator with partial grading for any step of the solution. Extending MATHEIS with this kind of possibility is our further occupation.

Sometime the learners need motivation. In this cases it is good to organize the tests in form of games, for example lottery game. This will make the process of learning more acceptable by the learners as fun. This is not offered by WIRIS plugin. Currently we are working on development of gamification strategies for improvement of this shortage.

Finally, it is good for the educator to use a question type that combines the features of "Fill in multiple blanks" and "Numerical answer" questions. Specifically, it is good to have a possibility for the fields in the "Fill in multiple blanks" question type to allow numerical answers. WIRIS quizzes do not possess this possibility.

## 7. CONCLUSIONS

The e-learning is a process. This process must be delegated by some management system – Learning Management System (LMS). E-content is a very powerful tool of education. E-content is valuable to the learners and also helpful to educators of all individual instruction systems.

Learning Management Systems have to be reliable and highly available. They must be secure and affordable. Such mission critical applications as LMSs require high performance load balancing.

By an analysis of few popular LMSs we have decided to install Sakai LMS in provision of load balancing and scaling. Sakai supports more math sciences e-content standards and tools than the other mentioned LMSs. MATHEIS agent based system for learning mathematics and informatics was embedded in the Sakai environment at the Faculty of Natural Sciences and Mathematics. The new developed GUPA agent is based on MathOpen and MathML standards and establishes communication with the mathematical WIRIS plugin.

WIRIS plugin for Sakai can help the users in easy manipulating the math symbols, formulas, equations, etc. in MATHEIS. Some problematic issues of the WIRIS usage like grading knowledge of math students by quizzes when the solutions process in math is based on many formulas, complex expressions and definitions have been noticed.

We have been considering further development of GUPA in order to give some additional intelligence in respect to semantic meaning and reasoning of the math sciences e-content and installed Sakai plugins.

Faculty of Natural Sciences and Mathematics, Department of Informatics, "Sts. Cyril and Methodius University", Skopje, Macedonia
*E-mail address*: mimozapetrova@yahoo.com.

Faculty of Natural Sciences and Mathematics, Department of Informatics, "Sts. Cyril and Methodius University", Skopje, Macedonia
*E-mail address*: cabukv@hotmail.com

Faculty of Natural Sciences and Mathematics, Department of Informatics, "Sts. Cyril and Methodius University", Skopje, Macedonia
*E-mail address*: roman.golubovski@t.mk